\documentclass[letterpaper, 10 pt, conference]{ieeeconf}
\usepackage{graphicx}      

\usepackage{xcolor}
\usepackage{amssymb}
\usepackage[ruled]{algorithm2e}  
\usepackage[htt]{hyphenat}
\usepackage{lipsum}
\usepackage{xspace}
\usepackage{soul}
\usepackage{mathtools}
\usepackage{makecell}
\usepackage{nicefrac}
\usepackage{float}
\usepackage{wrapfig}
\usepackage{ACSD-CR}

\newcommand{\bs}[1]{\ensuremath{\boldsymbol{#1}}}

\definecolor{dgreen}{rgb}{0.0, 0.5, 0.0}

\newcommand{\spc}{\ensuremath{\,\,}}	

\makeatletter
\newcommand{\subalign}[1]{%
	\vcenter{%
		\Let@ \restore@math@cr \default@tag
		\baselineskip\fontdimen10 \scriptfont\tw@
		\advance\baselineskip\fontdimen12 \scriptfont\tw@
		\lineskip\thr@@\fontdimen8 \scriptfont\thr@@
		\lineskiplimit\lineskip
		\ialign{\hfil$\m@th\scriptstyle##$&$\m@th\scriptstyle{}##$\crcr
			#1\crcr
		}%
	}
}

\newif\if@secthm
\if@secthm
\newtheorem{thm}{Theorem}[section]
\@addtoreset{thm}{section}
\else
\newtheorem{thm}{Theorem}
\fi

\newtheorem{defn}[thm]{Definition}

\newtheorem{rem}[thm]{Remark}

\IEEEoverridecommandlockouts   
\usepackage[T1]{fontenc} 
\usepackage[scaled=.7]{beramono}

\title {Extraction of a computer-certified ODE solver} 

\author{Grigory Devadze, Lars Flessing, Stefan Streif$^*$}

\usepackage{atbegshi}
\AtBeginDocument{\AtBeginShipoutNext{\AtBeginShipoutDiscard}}

\makeatletter
\let\NAT@parse\undefined
\makeatother
\usepackage[numbers]{natbib}
\begin{document}
\SETCR{\CRECC{}{}{}}

\thanks{* Technische Universitt Chemnitz,\\
	Automatic Control and System Dynamics Lab, Germany \\\{grigory.devadze,lars.flessing,stefan.streif\}@etit.tu-chemnitz.de}

\maketitle

\pagestyle{empty}

\begin{abstract}                
Reliably determining system trajectories is essential in many analysis and control design approaches.
To this end, an initial value problem has to be usually
solved via numerical algorithms which rely on a certain software realization.
Because software realizations can be error-prone, proof assistants may be used to verify the underlying mathematical concepts and corresponding algorithms.
In this work we present a computer-certified formalization of the solution of the initial value problem of ordinary differential equations.
The concepts are performed in the framework of constructive analysis and the proofs are written in the \texttt{Minlog} proof system.
We show the extraction of a program, which solves an ODE numerically and provide some possible optimization regarding the efficiency.
Finally, we provide numerical experiments to demonstrate how programs of a certain high level of abstraction can be obtained efficiently.
The presented concepts may also be viewed as a part of preliminary work for the development of formalized nonlinear control theory, hence offering the possibility of computer-assisted controller design and program extraction for the controller implementation.
\end{abstract}



\section{Introduction}

We consider the initial value problem
\begin{align}\label{eq:sys}
\dot x = f(t,x), \phantom{text} x(0)=x_0,
\end{align}
where $f:D \rightarrow \mathbb{R}^n$ is locally Lipschitz in $x$, continuous in $t$ and is defined on $D \subseteq \mathbb{R^+}\times \mathbb{R}^n$ with $(0,x_0) \in D$ .
It is well known that the ordinary differential equations (ODE) do not possess closed-form solutions in general.
Thus the initial value problem is an important subject in numerical system analysis and controller design.
Usually to apply a numerical method, one would need to predetermine the necessary parameters e.g. precision and step size which may be difficult to obtain beforehand in general.
Additionally, pure numerical approximations are not suitable for safety critical applications which require a certain level of rigor and correctness guarantees (see \cite{wichmann1992note} for general IEEE-754 floating point number issues or \cite{delmas2009towards,goodloe2013verification} for numerical issues which arise in avionics).

The gap between a mathematical concept and its implementation is sometimes also called \textit{algorithmic uncertainty}. The importance of deriving (as opposed to just implementing) correct precision-aware controllers within the so-called algorithmic control theory has been highlighted in \cite{Tsiotras2017-toward}.
Due to the mathematical nature of the problem, computer-assisted formal methods may be used to close this gap and to guarantee correctness fully \cite{hasan2015formal}. In this work we rely on the use of \textit{proof assistant} software, which have   been already successfully applied for important mathematical problems \cite{gonthier2013machine}. Moreover we are interested in the so-called \textit{proofs-as-programs paradigm} to investigate the relation of proofs with computer programming. This procedure is also called \textit{program extraction}, i.e. a certain proof of the existential theorem corresponds to a computer program which realizes the desired properties correctly. This paradigm can be especially interesting for controller design. Consider e.g., a controller implemented as a software program which has to guarantee a desired property such as stability of the closed-loop system. Note that even though development tools such as MATLAB possess certain possibilities for formal code verification \cite{lee2013requirements}, this does not guarantee that the implemented controller fulfills the desired properties because the controller itself did not go through a formal analysis. In this work we contribute towards formalized nonlinear control theory with automated program extraction. Please note that related works are reviewed in Section \ref{sec:rel_work}.
As a first step towards this goal, we focus on initial value problems for ODEs, which is undoubtedly a key concept in control theory.


\textbf{Contributions and outline of this work.}
The aim of the present work is to provide computer-assisted proofs for the initial value problem (\ref{eq:sys}).
As we are interested in the proof-as-programs paradigm, we obtain an extracted program, which solves the ODE numerically.
Our framework is carried out within a particular realization of $exact \spc real \spc arithmetic$: constructive analysis. The extracted program is still of a certain high level of abstraction, but can be also run as a usual numerical program.

We start with a review of related work on guaranteed and formal methods in Section \ref{sec:rel_work}.
The employed proof system and the preliminaries of formalization are explained in Section \ref{sec:prelim}.
The main result, i.e. the formalization of the Cauchy-Euler approximation is derived in Section \ref{sec:main}.
In Section \ref{sec:examples} we show the application of the derived program with some initial examples and compare the result with existing approaches in proof assistants.
In Section 5 we will give an outlook for how these concepts can be further used as a preliminaries for the development of formalized nonlinear control theory.
\section{Related works}\label{sec:rel_work}
To overcome the limitations of missing guarantees in numerical schemes, many different approaches for verified integration have been proposed.
\cite{Berz1998Verified} presented a method based on local modelling with high-order Taylor polynomials and a bound on the remainder term.
\cite{Rauh2011IntervalMethods} utilized the \texttt{ValEnclIA-IVP} to verify whether the enclosure of all reachable states remains within given bounds for known control strategies.
However these approaches are still numerical and although the used concepts are of a high abstraction level (e.g. Schrauder's fixed point theorem in (\cite{Berz1998Verified}), they do not have a formal connection to the actual programming code.
In the context of dynamical systems,  \cite{boldo2013wave} formally proves the correctness of a program implementing the numerical resolution of a wave equation, with Frama-C (\cite{cuoq2012frama}). In \cite{immler2019flow}, the authors formalize  variational equations related to the solutions of ordinary differential equations in Isabelle/HOL \cite{Nipkow02IsabelleHol}.
With respect to the initial value problem the work of \cite{Immler2018-verified-ode} provides the verified implementation of an approximation of a solution of an ODE based on arbitrary precision floating-point numbers.
In the present work, the implementation is done by the accurate representation of special infinite sequences of rational numbers.
This allows a parametrization of the algorithm by the demanded error bound with the guarantee that this bound holds.
Similarly it was done in \cite{Makarov2013-picard} where the algorithm is derived from the Picard-Lindel\"of theorem.
However, the level of abstraction of the theorem is very high, such that the running times of the algorithm on even small instances are unusable in practice.
In the current work, the level of abstraction is reduced and the approximate solution of the initial value problem is derived efficiently in a direct way by means of the program extraction techniques.
Additionally the correctness of the programs follows immediately given from the Soundness Theorem \cite{Berger2011-minlog}.

Proof assistants also play an important role in formal verification of control systems. \cite{Zou2013-form-ver-Chinese-train} used Isabelle for implementation of their control system formal verification based on Hybrid Hoare logic. The Why3 platform \cite{bobot2011why3}  coupled with MATLAB/Simulink was used in \cite{Araiza-Illan2014-thm-prover-sys, Araiza2015-thm-proving-Simulink} to perform simple stability checks of linear discrete systems with quadratic Lyapunov functions. \cite{Gao2014-descr-ctrl-thr} mentioned several proof assistant software tools to be considered for implementation of formal verification of control systems, including \cite{CoqProofAssistant}, HOL-Light \cite{harrison1996hol}, Isabelle/HOL and Lean \cite{de2015lean}. A closely related work is due to \cite{Anand2015-roscoq} where the authors present the framework for the certified programs for robots based on constructive real numbers \cite{Oconnor2008-certified}.
The important work by \cite{Platzer2012-dyn-sys-log} suggested the first-order formal system for the formalization of hybrid systems, the so-called differential dynamic logic (\texttt{dL}).
The key tool which lies in the heart of this formal system is quantifier elimination on the real closed fields \cite{Tarski1998-quantifier-elim}. Notably, \cite{bohrer2017formally} uses classical formalizations of Picard-Lindel\"of theorem to provide soundness results for important \texttt{dL} axioms: Differential Solution (\texttt{DS}) and Differential Ghost (\texttt{DG}).

\if{
{\color{blue} 
Formally Verified Differential Dynamic Logic \cite{bohrer2017formally}.\\
Formalization of soundness of \(\delta L\).\\
Logic for hybrid systems, ``cyber-physical systems''.\\
Soundness proof for KaYmaera logic core by transporting the core properties and reasoning in Coq and Isabelle.\\
Extend for verified prover kernels in Isabelle and Coq.\\
\begin{itemize}
	\item Constant Differential Solve DS states that constant ODEs are uniquely solved by linear functions\\
	\([x'=f\&q(x)] \leftrightarrow \forall t \geq 0 \left(\left(\forall 0 \leq s \leq t q(x+fs)\right)\rightarrow [x:=x+ft]p(x)\right)\)\\
	\begin{itemize}
		\item \(\forall c, x'=c \rightarrow x(t) = x_0+ct\)\\
		A constant differential equation has a linear function as solution.
		\item Proof in isabelle by Picard-Lindelöf.
		\item Results in any local Lipschitz ODE has unique solution called ``flow''
		\item Prove by showing that \(f(t) = x_0+ct = flow\) 
		\item proof in Coq \(c,r \in \mathbb{R}, g:=\mathbb{R}\rightarrow\mathbb{R}, \forall x\leq r, g\in \mathcal{C} \wedge \frac{\mathrm{d}g}{\mathrm{d}x=c} \rightarrow g(r)=g_0+cr\)
		\item Direct proof of uniqueness by showing for all linear functions, the derivative is a constant.
		\item Basic rules\dots
		\item No actual implementation of a accurate solution, but a trivial case to construct more complex form
	\end{itemize}
	\item Differential Ghost DG states that	one can add equations to ODEs as long as they are linear,	implying they have solutions of the same duration.\\
	\([x'=f(x)\&q(x)]p(x) \leftrightarrow \exists y[x'=f(x),y'=a(x)y+b(x)\&q(x)]p(x)\)\\
	\begin{itemize}
		\item Adding a linear ODE to another ODE\\
		\(\Phi_x := x'= x^2\) extend to \(y'=x^3y+2y\). With \(\Phi_y(\Phi_x,y) := y(0)=y_0, y'(t)=\Phi_x(t)^3y+2y\).\\
		\item linear ODE, and this has a solution.
		\item Proven by reducing the intervall to \([0,t]\) for \(\Phi_x\) and \(\Phi_y\), prove locally Lipschitz.
		\item This is done by showing that \(\Phi_y\) has a constant derivative, and then prove that \(\Phi_x\) is bounded by a linear function. By working on a metric functional space with norm \(\vert\vert f\vert\vert = \underset{x\neq 0}{\mathrm{sup}}\frac{\vert\vert f(x)\vert\vert}{\vert\vert x \vert\vert}\) continuity needs to be shown for each component separately.
		\item In Coq only ``Production rule'' is proven \(\rightarrow\).	
	\end{itemize}
\end{itemize}

\cite{platzer2017complete}
This is the lone definition of the \(dL\) from Platzer. This is the ``Draft'' to the upper one

\cite{Rauh2011IntervalMethods}
Interval Methods for Verification and Implementation of Robust Controllers
\begin{itemize}
	\item Utilizing the ValEncIA-IVP lib (initial value problems)
	\item Iterative mehtod with upper error bound
	\item Used for uncertain initial values
	\item Lib can be use to prove whenether the enclosure of all reachable states remain within given bounds for know control strategies.
	\item It can verify in a feedback system that the controller !! ``Matches the corresponding physical input constraints''
	\item How to use verified solutions of IVP for diffrerential algebraic expressions
	\item use DAEs for verified Feedforward control and state estimations
	\item Last point avoids the point of simulation or verification of control strategies.
	\item Basically this setting is to employ the toolbox for constraint propagation.
	\item This is in the a full numerical approach relying on the solution of the IVP.
\end{itemize}

\cite{Berz1998Verified}
Straight forward. The application of the fixed point theorem makes this not verifiable, since not computable, since no unique solution must exist.
Due to reduction of a differential equation to an operator a fixed-point problem arrises to solve the equation \(\mathbf{r} = A(\mathbf{r})\).

\cite{platzer2008keymaera}
\begin{itemize}
	\item Example of how KeYmaera is used.
	\item Verbal explanation of the conversion of the systems
	\item Three staged\\
	\begin{enumerate}
		\item Convert to Polynomial function if fails,
		\item Convert to inductive differentials if fails,
		\item introduce a value \(x \in \mathbb{R}\) that represents the solution of the ODE
	\end{enumerate}
	\item Verification of the whole problem is within the differential Logic but
	\item For the introduced real values it is over the domain \(\mathbb{R}\)
	\item To ensure correctness and the inablility to convert the functions into the system the claim needs to hold \(\forall x \in \mathbb{R}\)
	\item Final verification might still depends on solution of the ODE.
	\item Might provable restrictions of the ODE useable for a tightening of the domain \(\forall x \in D \subset \mathbb{R}\)
	\item Currently handled by ``Real elimination''
	\item Find reals for that the constraints hold and gain properties that the ODE need to fullfill.
\end{itemize}

\cite{tsiotras2017toward}
\begin{itemize}
	\item Introducing Algorithmic control.
	\item More focused on algorithmic control of different control strategies.
	\item Similar to MPC/RHC.
	\item Setup an controller to be an algorithm controlling the system and the control law.
	\item Main perspective are on feasible solutions anyway
	\item Basic task to incorporate discrete and continuous system into single framework, hybrid systems
	\item Rise of extended questions like computability, decidability, complexity
	\item necessary to prove implementations, beyond mere mathematics.
\end{itemize}
}
}
\fi
\section{Minlog Proof System} \label{sec:prelim}
In this section we introduce some preliminary concepts of the proof system \texttt{Minlog}, the notion of totality and the concept of inductive definitions. Afterwards we give the details how the formalization works in constructive setting.
\subsection{Neccessary preliminaries}
In the present work, the analysis and formalization are done within the framework of constructive analysis with a special focus on a concrete and efficient computational realization as well as formal verification.
	
	
	

The interactive proof system \texttt{Minlog} has been developed with the special intention of the program extraction from mathematical proofs.
The atomic constructions are done by introducing of free algebras as types which are given by its constructors. 
The computation is implemented efficiently by the usage of the so-called normalization-by-evaluation.
The reason of usage of typed theories is that the normalization-by-evaluation technique, which is applied to the proof terms, needs typed formulas.
As an example consider the construction of the type \texttt{nat} of natural numbers $\mathbf{N}$ by stating that \texttt{Zero} is \texttt{nat} and \texttt{Succ} is a mapping of type \texttt{nat$\Rightarrow$nat}.
The normalization of the term 1+2 would result in \texttt{Succ(Succ(Succ Zero))}.

Program extraction is done from the constructive proofs of theorems.
Additionally a proof that an extracted program realizes its specification is machine generated.
The  user-defined computation and corresponding rewrite rules are assigned to the so-called program constants.
Usually the program constants are recursively written typed programs.
Consider for example the multiplication of two natural numbers.
The corresponding program constant \texttt{NatTimes} is of type \texttt{nat $\Rightarrow$ nat $\Rightarrow$ nat} and the computational rules are given recursively as:
\begin{align*}
	&\texttt{NatTimes}\spc  n \spc \texttt{Zero} \rightarrow  \texttt{Zero} \\
	&\texttt{NatTimes} \spc n \spc \texttt{Succ} \spc m \rightarrow \texttt{NatTimes}\spc n \spc m + n
\end{align*}

\subsection{Inductive definitions and theorem proving}
Before one starts to prove theorem with the help of the computer, one has to state some definitions which are relevant for the problem statement.
In \texttt{Minlog} it is done by the introducing the so-called inductively defined predicate constant (IDPC).
Informally speaking, these are formations of implications which leads to the proposition which has to be defined.

As an example assume that one wants to provide a definition for rational numbers which lie in an unit interval.
Then the corresponding IDPC \texttt{RatInUnitI} is given by:
\begin{equation*}
	\forall a \in \mathbf{Q} \spc 0 \le a \rightarrow a \le 1 \rightarrow \texttt{RatInUnitI} \spc a
\end{equation*}
which means semantically that to show $ \texttt{RatInUnitI} \spc a$ one has to show that $0 \le a$ and $a \le 1$.

\subsection{Preliminaries for the formalization in constructive setting.}
This part is based on the foundation of constructive analysis with witnesses \cite{Schwichtenberg2006-constructive}. 
The most important object in the presented setting is a real number.
In \texttt{Minlog} reals are build by introducing the typed variables \texttt{as:$\mathbf{N}\Rightarrow\mathbf{Q}$}, \texttt{M:$\mathbf{P}\Rightarrow\mathbf{N}$}, defining the constructor \texttt{RealConstr:$(\mathbf{N}\Rightarrow\mathbf{Q})\Rightarrow(\mathbf{P}\Rightarrow\mathbf{N})\Rightarrow\mathbf{R}$} and introducing the IDPC \texttt{Real} i.e. the real $x$ is given by a regular Cauchy sequence of rationals $(a_n)_n$ with a given weakly increasing modulus $M$.
That is $\forall p,m,n \spc |a_{n} - a_{m}|\le 2^{-p}$ whenever $M(p) \le n,m$ and $\forall p,q \spc p \le q \rightarrow M(p) \le M(q)$.

Two reals $x=((a_n)_n,M)$ and $y=((b_n)_n,B)$ are equal, if $\forall p\in \mathbf{P}\spc|a_{M(p+1)}-b_{N(p+1)}|\le 2^{-p}$.
A real $x$ is called non-negative if $\forall p \in \mathbf{P} \spc 2^{-p} \le (a_n)_{M(p)}$.
For a real $x$ denote $x \in_p \mathbf{R}^+ $ (alternately $0<_p x$) if $2^{-p} \le (a_n)_{M(p+1)}$ for some $p$.
\begin{defn}[\texttt{Cont}]
	An uniformly continuous function $f:I \rightarrow \mathbb{R}$ on a compact interval $I$ with rational endpoints is given by
	\begin{enumerate}
		\item an approximating map $h_f(I \cap \mathbf{Q} ) \times \mathbf{N} \rightarrow \mathbf{Q} $ and a map $\alpha_f: \mathbf{P} \rightarrow \mathbf{N}$ such that $(h_f(a_n,n))_n$ is a Cauchy sequence with modulus $\alpha_f$
		\item a modulus $\omega_f:\mathbf{P} \rightarrow \mathbf{P}$ of continuity, which satisfies 
		\begin{equation*}
		|a-b| \le 2^{-\omega_f(p) +1} \rightarrow |h_f(a,n)-h_f(b,n)|\le 2^{p}
		\end{equation*}		
	\end{enumerate}
	with $\alpha_f$ and $\omega_f$ weakly increasing.
\end{defn}

The equivalent realisation of continuous functions in \texttt{Minlog} is done via introducing of the type \texttt{cont} and the IDPC \texttt{Cont f}, stating that $f$ is a proper continuous function.
Firstly the basic constructors of \texttt{cont} are:
\begin{itemize}
	\item $h$: $\mathbf{Q}\Rightarrow\mathbf{N}\rightarrow\mathbf{Q}$ (approximation mapping)
	\item $\alpha$: $\mathbf{P}\Rightarrow\mathbf{N}$ (modulus of convergence)
	\item $\omega$: $\mathbf{P}\Rightarrow\mathbf{P}$ (modulus of continuity)
\end{itemize}
and finally the constructor \texttt{ContConstr: $\mathbf{Q}\Rightarrow \mathbf{Q}\Rightarrow h\Rightarrow\alpha\Rightarrow\omega\Rightarrow$ cont}.
The property that the mapping of type \texttt{cont} is truly a continuous function in the presented sense can be encoded via the following IDPC \texttt{Cont}:\\
\texttt{$\forall a0,b0,h,\alpha,\omega \spc$}:\\
\texttt{$(\forall a(a0\le a \rightarrow a \le b0 \rightarrow$ Cauchy$(h \spc a)\spc\alpha))$}$\rightarrow$\\
\texttt{$(\forall a,b,k,n(a0\le a \rightarrow a \le b0 \rightarrow a0\le b \rightarrow b \le b0 \rightarrow \alpha\spc p \le n \rightarrow |a-b|\le 2^{\omega \spc p})\rightarrow |h \spc a - h \spc b| \le 2^{-p})$}$\rightarrow$\\
\texttt{$\forall p,q (p\le q \rightarrow \alpha \spc p \le \alpha \spc q)$}$\rightarrow$\texttt{$\forall p,q (p\le q \rightarrow \omega \spc p \le \omega \spc q)$}$\rightarrow$\texttt{Cont(ContConstr a0 b0 $h$ $\alpha$ $\omega$)} \\
As an example consider the representation of the function $f(x)=x^2$ on the interval [0,2]:\\
\texttt{ContConstr 0 2 ([a,n]a*a) ([p] 0) ([p] p+3)}\\
\if{
Similarly a function on multiple variables can be defined as follows:
\begin{defn}
	An uniformly continuous function $f:B \rightarrow \mathbb{R}$ on a compact ball $B=(c,R) \subset \mathbf{R}^n$ with rational center $\bs c$ and rational radius $R$ is given by
	\begin{enumerate}
		\item an approximating map $h_f(B \cap \mathbf{Q}^n ) \times \mathbb{N} \rightarrow \mathbb{Q} $ and a map $\alpha_f: \mathbb{P} \rightarrow \mathbb{N}$ such that $(h_f(\bs a_n,n))_n$ is a Cauchy sequence with modulus $\alpha_f$
		\item a modulus $\omega_f:\mathbb{P} \rightarrow \mathbb{P}$ of continuity, which satisfies 
		\begin{equation*}
		\|a-b\|_1 \le 2^{-\omega_f(p) +1} \rightarrow |h_f(a,n)-h_f(b,n)|\le 2^{p}
		\end{equation*}		
	\end{enumerate}
	with $\alpha_f$ and $\omega_f$ weakly increasing.
\end{defn}}\fi
The definition of the vector-valued function is provided similarly:
\begin{defn}[\texttt{ContVec}] \label{def:contvec}
	An uniformly continuous vector-valued function $\bs f:B \rightarrow \mathbf{R}^n$ on a compact ball $B(c,R) \subset \mathbf{R}^n$ with rational center $\bs c$ and rational radius $R$ is given by
	\begin{enumerate}
		\item an approximating map $\bs h_f(B \cap \mathbf Q ^n ) \times \mathbf{N} \rightarrow \mathbf{Q} $ and a map $\alpha_f: \mathbf{P} \rightarrow \mathbf{N}$ such that $(h_f(a_n,n))_n$ is a Cauchy sequence with modulus $\alpha_f$
		\item a modulus $\omega_f:\mathbf{P} \rightarrow \mathbf{P}$ of continuity, which satisfies 
		\begin{equation*}
		\|\bs a-\bs b\|_1 \le 2^{-\omega_f(p) +1} \rightarrow \|\bs h_f( \bs a,n)- \bs h_f( \bs b,n)\|_1\le 2^{p}
		\end{equation*}		
	\end{enumerate}
	with $\alpha_f$ and $\omega_f$ weakly increasing.
\end{defn}
\begin{rem}
	The choice of the norm $\| \cdot \|$ influences the values of moduli $\alpha_{\bs f}$ and $\omega_{\bs f}$.
Usually, one would prefer to use either $1$-norm or $\infty$-norm for  computational purpose, since for rational vectors the values of the norm are also rationals, which is not the case, e.g., for the Euclidean norm.
\end{rem}
\begin{defn}[\texttt{Application}]
	The application of an uniformly continuous function $f:I\rightarrow \mathbb{R}$ to a real $x:=((a_n)_n,M))$ in $I$ is defined to be a Cauchy sequence
	$h_f(a_n,n)_n$ with modulus $\max(\alpha_f(p+2),M(\omega_f(p+1)-1).$ We denote this real by $f(x)$.
\end{defn}
The following theorem is necessary to perform computations with the real numbers.
\begin{thm}[\texttt{RealApprox}]
	$\forall x,p \spc \exists a (|a-x|\le 2^{-p})$.
\end{thm}
At this point an easy example of the program extraction in \texttt{Minlog} is provided.
Consider the following goal as an equivalent to the theorem statement above:
\begin{center}
	\texttt{$\forall$ x,p Real x $\rightarrow$ $\exists$ a abs(a-x)$\le$(1\#2)**p }
\end{center}
After unfolding the definition of $x=((a_n)_n,M)$ and assuming universally quantified variables $(a_n)_n, M$ and the hypothesis \texttt{Real x} the new goal is
\begin{center}
	\texttt{$\exists$ a abs(a-(RealConstr $ (a_n)_n\spc M$))$\le$(1\#2)**p}.
\end{center}
Here the existential quantifier can be eliminated by the term \texttt{(ex-intro (pt "$(a_n)_{M(p)}$")}.
The rest of the proof uses usual techniques to prove that the $|(a_n)_{M(p)} - ((a_n)_n,M)| \le 2^{p}$.
Once the existential proof of the theorem is done, this theorem can be animated.
In case of the animation of  \texttt{RealApprox} one obtains a program constant with the following computation rule:	

\texttt{cRealApprox $\rightarrow$ 
		[x0,p1]}\\
\texttt{\phantom{text}[if x0 ([as2,M3]as2(M3 p1))]}
which is an extracted program from the constructive proof of the theorem \texttt{RealApprox}.
Moreover, one can use the extracted program in the proof of the following theorem and generate new program:
\begin{thm}[\texttt{RealVecApprox}]
	$\forall \bs x \in \mathbf{R}^n ,p \spc \exists \spc \bs e \in \mathbf{Q}^n \spc \|\bs e - \bs x \|_1 \le 2^{-p}$.
\end{thm} 
\begin{proof}
	Proceed by induction on n.
	For $n=1$ the existence holds by the theorem RealApprox and is realized by \texttt{cRealApprox}.
For $n=2$ and for any vector real number $\bs x = (x_1,x_2)$ generate a vector $\bs e=(e_1,e_2)$  with $|e_1-x_1| \le 2^{-(p+1)},|e_2-x_2| \le 2^{-(p+1)}$ again by the Theorem RealApprox.
Clearly $|e_1-x_1| + |e_2-x_2| \le 2^{-p}$.
For the induction step assume that for the vector $\bs x'=(x'_1,...,x'_n)$ there exists a vector $\bs e'$ with $\| \bs e' - \bs x'\| \le 2^{-(p+1)}$ and for the real number $x_{n+1}$ there exists a rational number $e_{n+1}$ again by RealApprox such that $|e_{n+1} - x_{n+1}|\le 2^{-(p+1)}$.
Then for the vector $\bs x = (x_1,...,x_{n+1})$ the vector $e=(e',...,e_{n+1})$ satisfies the desired condition and is a $2^{-p}$ approximation to a real vector $x$.
\end{proof}

Thus we are able to perform correct computations with vector-valued functions, since the program constant \texttt{Application} is capable to generate new real number vectors $\bs f (\bs x)$ correctly.

\begin{defn}[Time derivative]
	Let $\bs \varphi_1:[T_1,T_2] \rightarrow \mathbf{R}^n ,\bs \varphi_2:[T_1,T_2] \times B \rightarrow \mathbf{R}^n$ be continuous in sense of Definition \ref{def:contvec} and $0 \le T_1 < T_2$ .
The function $\bs \varphi_2$ is called partial derivative w.r.t. the time of $\bs \varphi_1$ with the modulus of differentiability $\delta_{\bs \varphi_1}:\mathbf{P} \rightarrow \mathbf{P}$ if for $t_1,t_2 \in [T_1,T_2]$ with $t_1<t_2$,
	\begin{align*}
	&\forall p \spc t_2\le t_1+2^{- \delta_{\bs \varphi_1(p)}} \rightarrow\\&|\bs \varphi_1(t_2)-\bs \varphi_1(t_1)-\bs \varphi_2(t_1,\bs x)(t_2-t_1)|\le 2^{-p}(t_2-t_1)
	\end{align*}
\end{defn}
We denote the function $\bs \varphi_2$ as $\dot {\bs \varphi_1}$ and call the function $\bs \varphi_1$ differentiable on $[T_1,T_2]$.

The next steps allow to generate a uniform partition of a rational interval such that the functions can be evaluated at discrete values.
\begin{defn}[\texttt{Partition}]
	Let $a,b$ be rational numbers such that $a<b$.
A list $P=a_0, \dots, a_n$ of rationals is a partition of the interval $[a,b]$, if $a=a_0 \le a_1 \le \dots \le a_n=b$.
\end{defn}
\if{
The corresponding IDPC is:\\
\texttt{Partition}:\\
\texttt{$\forall a,b,c,$als}:\\
\texttt{$(a<b \rightarrow$ Head(als)==$a$ $\rightarrow$ Last(als)==$b$
$\rightarrow$ $1<$Lh als) $\rightarrow$}
\texttt{$\forall n(n \le$ Pred(Pred(Lh als)) $\rightarrow$ (als\_\_(Succ $n$)$+$ ~(als\_\_$n$)) $\le$ abs($c$)) $\rightarrow$ PartitionIntro} 
}
\fi

\begin{thm}[\texttt{RatLeAbsBound}]
	$\forall a \in \mathbf{Q} \spc \exists p \in \mathbf{P} \spc |a| \le 2^{p}.$
\end{thm}
The corresponding extracted program of type $\texttt{rat} \Rightarrow \texttt{pos}$ is denoted as \texttt{cRatLeAbsBound}.

\begin{defn}
	The program constant \texttt{UnifP} of type \texttt{nat$\Rightarrow$} \\ \texttt{rat$\Rightarrow$rat $\Rightarrow$(list rat)} is recursively defined as follows:
	\begin{itemize}
		\item \texttt{UnifP Zero a b} $\rightarrow$ \texttt{b:}
		\item \texttt{UnifP (Succ n) a b} $\rightarrow$\\ \texttt{UnifP (n) a ((a+b)/2)} :+: \texttt{UnifP (n) a ((a+b)/2)}
	\end{itemize}
where \texttt{:+:} is list concatenation.
\end{defn}

\begin{defn}
	The program constant \texttt{UnifP} of type \texttt{rat$\Rightarrow$}\\ \texttt{pos$\Rightarrow$ (list rat)} is  defined as:\\
	\texttt{(0)::UnifP Succ(p+cRatLeAbsBoundPos a) 0 a}
\end{defn}
The direct statement of the goal 
\begin{align*}
	\forall b,p \spc \texttt{Partition} \spc 0 \spc b \spc 2^{-p} \spc \texttt{UnifP b p }
\end{align*}
corresponds to the statement of the following theorem.
Thus we are able to prove that the program constant \texttt{UnifP} indeed generates a list which is \texttt{Partition}.
\begin{thm}[\texttt{UnifPPartition}]
	For $a \in \mathbf{Q}, p \in \mathbf{P}$  the list of rationals generated by the program constant \texttt{UnifP} is a partition $P=c_0,...,c_n$ of the interval $[0,a]$.
Additionally it holds that $\max \{ c_{i+1} - c_i  | i<n \} \le 2^{-p}$.
\end{thm}
\begin{defn}[\texttt{EulerMap}]
	The program constant \texttt{EulerMap} of type \texttt{contVec=>ratVec=>nat=>rat=>pos=>ratVec} is recursively defined as follows:
	\begin{itemize}
		\item \texttt{EulerMap} $\bs f$ $ \bs a$ $ \texttt{Zero} \spc \texttt{d} \spc \texttt{p} \rightarrow \bs a$  
		\item \texttt{EulerMap} $\bs f$ $ \bs a$ $ \texttt{(Succ n)} \spc \texttt{d} \spc \texttt{p} \rightarrow$ \texttt{EulerMap} $\bs f$ $ \bs a$ $ \texttt{n} \spc \texttt{d} \spc \texttt{p} \\$
		$+ \texttt{d} \cdot $ \texttt{cRealVecApprox} ($\bs f$ (\texttt{n*d} ,\texttt{EulerMap} $\bs f$ $ \bs a$ $ \texttt{n} \spc \texttt{d} \spc \texttt{p})) \spc \texttt{p}$
	\end{itemize}
\end{defn}

In summary, in this section we presented the most important concepts of the \texttt{Minlog} system and presented the formalization of constructive analysis in \texttt{Minlog} as well as the extension of the standard library which was necessary for our goal.
In the next section we provide the formalization of Cauchy-Euler approximation method in our framework.


\section{Main Result} \label{sec:main}
\subsection{Formalized proof in \texttt{Minlog}}
Using the machinery presented in Section \ref{sec:prelim} we prove the existence of an approximate solution by the means of the Cauchy-Euler approximation method.
We consider a differential equation (properly stated in our constructive framework):
\begin{align} \label{sys:eq:prop}
\dot {\bs x} = \bs f(t,\bs x) , \phantom{text} \bs{x}(0)=\bs{x}_0.
\end{align}
Although the proof is standard, we provide it here in formalized way which is very close to a computer proof.

\begin{thm}[\texttt{CauchyEulerApproxSolCorr}] \label{thm:cauchyeuler}
Let $\bs f:B \rightarrow \mathbf{R}^n$ with $B(c,R) \subset \mathbf{Q}^n$  be an uniformly continuous vector-valued function and $(0,\bs x_0) \in B$.
Assume that for the set $B'$ given by $|t| \le t_a, \| \bs x - \bs{x_0} \|_1 \le x_b$ is in $B$.
Assume there exists a rational number $C>0$ such that $\|\bs f (t,\bs x) \|_1 \le C$ for $(t,\bs x) \in B'$.
Then for every $p \in \mathbf{P}$ there exist an approximate solution  $\bs \varphi:[0,T] \rightarrow \mathbf{Q}^n$ up to the error $2^{-p}$ in sense that
\begin{align*}
		\| \dot {\bs \varphi}(t) - \bs f(t,\bs \varphi(t))\|_1 \le 2^{-p}
\end{align*} for all $t \in [0,T]$ with $T=\min \left\{ t_a ,\frac{x_b}{M}\right\}$ where $\dot {\bs \varphi}(t)$ is defined.
\end{thm}
\begin{proof}
	For any $p \in \mathbf{P}$ obtain a number $q \in \mathbf{P}$ by the continuity such that:
	\[\|f( t_1, \bs x) - f(t_2,\bs y) \|_1 \le 2^{-(p+1)},\]
	whenever $\| \bs x - \bs y\|_1,|t_1-t_2| \le 2^{-q}$.
Generate a partition $P=c_0, \dots , c_n$ by applying the program constant \texttt{UnifP} of the interval $[0,T]$ such that $\max \{c_{i+1}-c_{i}| i<n\} \le \min \left( 2^{-q}, 2^{-(q + \texttt{cRatLeAbsBound } C)}\right)$.\\
Recursively construct the approximate solution $\bs {\varphi}$ via the program constant \texttt{EulerMap}.
Note that the approximate solution satisfies for $i=1, \dots , n$:
	\begin{align*}
		&\bs \varphi(t) = \bs b_i + (t - c_i) \bs s_{i-1}  \text{for } {c_{i-1}} \le t \le c_i\\
		&\bs b_i = \bs \varphi(c_i),
	\end{align*}
where $\bs s_{i-1}$ are calculated such that $\|\bs s_{i}\|_1 \le C$ and $\| s_{i-1}-f(\bs b_i,c_i)\| \le 2^{-(p+1)}$ by the program constant \texttt{cRealVecApprox}.\\
Observe that for $c_{i-1} < t < c_{i}$ 
\begin{align*}
	&\| \bs \varphi(t) - \bs b_{i-1} \|_1 \le |t - c_{i-1}|\|s_{i-1}\|_1\\ &\le 2^{-(q+\texttt{cRatLeAbsBound } C)} 2^{\texttt{cRatLeAbsBound } C)}\\
	&\le 2^{-q}
\end{align*}
and finally,
\begin{align*}
	 &\| \dot{\bs \varphi}(t) - f(t,\bs \varphi(t)	\|_1 = \| s_{i-1} - f(t,\bs \varphi(t))\|_1\\
	&\le \| s_{i-1} - f(c_i,\bs b_i) \|_1 + \|f(c_i,\bs b_i) - f(t,\bs \varphi(t)	\|_1 \\ & \le 2^{-(p+1)} + 2^{-(p+1)} \le 2^{-p}.
\end{align*}
\end{proof}

\begin{rem}
For existence and uniqueness we rely on the pen-and-paper version \cite{hurewicz1990lectures} and admit its correctness for now.
Let $\bs f: B \rightarrow \mathbb{R}$ be continuous and satisfy the Lipshitz condition w.r.t. its second argument, i.e. there exists a constant $L$ such that $ \|f(t,\mathbf{x}_1) -   \mathbf{x}_2)\| \le L \|\mathbf{x}_1 - \mathbf{x}_2\|$ for any $x_1,x_2 \in B$.
Assume that the assumption of the Theorem \ref{thm:cauchyeuler} holds.
Then we can construct a unique exact solution $\bs \psi:[0,T] \rightarrow \mathbb{R}^n$ to (\ref{sys:eq:prop}) such that $\bs {\psi}(0)=\bs x_0$.
Moreover let $\bs \phi:[0,T] \rightarrow \mathbb{R}^n$ be an approximate solution to (\ref{sys:eq:prop}).
Then \begin{align*}
	\forall p, t \in [0,T] \spc \| \bs \varphi (t) - \bs \psi(t) \|_1 \le \frac{2^{-p}}{L}\left(e^{L \cdot T-1}\right).
\end{align*}
\end{rem}
Thus proper evaluation of approximate solution $\bs \varphi$ yields arbitrary approximation of an exact solution of (\ref{sys:eq:prop}).
\begin{rem}
The extracted program $\bs \varphi$ is of type \texttt{contVec} and fulfills the assumptions of Definition \ref{def:contvec} (\texttt{ContVec}). The respective modulus of continuity $\omega$ and modulus of convergence $\alpha$ are derived by the proof system automatically. 
Moreover $\bs \varphi(t)$ is admissible, i.e. $(t,\bs \varphi(t)) \in B$ and piecewise continuous differentiable.
\end{rem}

\subsection{Program extraction}
The direct extraction yields the invocation of the program constant \texttt{EulerMap}, which is not usable for practical computation.
Thus we define new program constant \texttt{EulerMapFast} of the type \texttt{contVec=>(ratVec pair ratVec)=>rat=>pos=>(ratVec pair ratVec)} recursively as:
\begin{itemize}
	\item \texttt{EulerMapFast Zero rt d p} $\rightarrow$ \texttt{rt}   
	\item \texttt{EulerMapFast (Succ n) rt d p} $\rightarrow$ \texttt{EulerMapFast n ((rht rt) pair} \\
	\texttt{(rht rt + d*cRealVecApprox f(n*d,(rht rt)) p)  c p}
\end{itemize}
where \texttt{pair} is a constructor for a tuple, \texttt{rt} is a varible of type \texttt{(ratVec pair ratVec)} and \texttt{rht} accesses the right element of the tuple.

Next we showed the equivalence of outputs of the program constants \texttt{EulerMap} $\bs f$ $ \bs a$ $ \texttt{n} \spc \texttt{d} \spc \texttt{p}$ and \texttt{EulerMapFast n (Inhab rat) pair $\bs a$ $\texttt{d} \spc \texttt{p}$}, where \texttt{(Inhab ratVec)} is an arbitrary inhabitant of rational vectors.
Observe that \texttt{EulerMapFast} mimics a usual \texttt{for} loop and there are no more redundant recursive calls as in \texttt{EulerMap}.
However \texttt{EulerMap} yields easier goals for the reasoning i.e. we use \texttt{EulerMapFast} for a computational version of the Theorem \ref{thm:cauchyeuler} where the program equivalence guarantees the correctness.
Additionally we evaluate all elementary operations (+ , - , $\cdot$ , $/$, \texttt{floor}) on rational numbers externally in \texttt{Scheme} due to the fact that the operations which are defined for decidable/discrete predicates (rational numbers in this case) should be not affected by some algorithmic uncertainty.
Furthermore, \texttt{Scheme} cancels the greatest common divisor automatically, thus we do not have to be concerned about it during the theorem proving.
As for the last optimization we implemented the so-called compression of real numbers.
That is, for any real $x=((a_n)_n,M)$ there is a new real $y=((b_n)_n,p+2)$ with $(b_n)_n:=\frac{\texttt{floor}(a_{M(p)}\cdot 2^n)}{2^n}$ with $x=y$.
This allows us to control the length of the denominator and makes the Cauchy sequences less complex.
Applying this optimization allows performing the computation more efficiently which is shown experimentally in the next section.

\section{Case Study}\label{sec:examples}
In the following we compare our results with the existing approaches in proof assistants on the following basic but frequently used examples in this line of research \cite{Makarov2013-picard,Berz1998Verified}. 

\textit{Example 1:} 

For the differential equation 
\[
\dot x=x, x(0)=1
\] 
the data required for the application of Theorem \ref{thm:cauchyeuler} is: 
\begin{itemize}
	\item $h_{f}(a_1,a_2,n)=a_1$,
	\item $\alpha(p)=0$,
	\item $\omega(p)=p$.
	\item $C=2$, $L=1$.
\end{itemize}
We obtain two correct decimal digits ($p=10$) at $t=\frac{1}{2}$ i.e.
the result is the rational number $\frac{55317227}{33554432}$ within 3 sec, whereas Coq C-CORN needs approximately 10 min.
As for the work of \cite{Immler2018-verified-ode}, the timing is comparable, however, we did not need to consider the notion of the stability of the Euler method, since our approach runs an "ideal" Euler method efficiently.
Moreover, we stay within \texttt{Minlog}'s verification kernel except for some basic computations on rationals.

\textit{Example 2: Integrating the circle} 
\begin{align*}
	\dot x = -y, & \spc \spc x(0)=1\\
	\dot y = x, & \spc \spc y(0)=0
\end{align*}
We use the following representation:
\begin{itemize}
	\item $\bs h_{\bs f}(a_1,a_2,a_3,n)=(-a_2,a_1)$,
	\item $\alpha(p)=0$,
	\item $\omega(p)=p$,
	\item $C=2$, $L=1$
\end{itemize}
For two correct decimal digits of the integration from 0 to 2$\pi$ the resulting rational vector is $\left(\frac{16790121}{16777216}, \frac{22267}{134217728} \right)$ and its timing is approximately 30 min which corresponds to approximately 26 000 verified evaluations of $\bs f$.
Note that our assumptions are valid for any $t$, thus we are allowed to run the algorithm successively and obtain the continuation of the solution.

Overall we conclude that our extracted program achieved relatively good performance. 
Additionally, observe that we are able to obtain a verified result at any desired precision in principle. Please note that extracted programs in \texttt{Minlog} allow the evaluation of verified high-level functions directly. Clearly the procedure of program extraction is not limited to the optimization presented in Section \ref{sec:main} and the running time can be further reduced drastically. A detailed analysis is beyond the scope of this work and subject of current investigations. Note further that while the analysis of trajectories is often essential for the controller design and its formal verification, the controller acting online may be much simpler and may not need to determine the trajectory.

\section{Conclusion}
We formalized the Cauchy-Euler approximation method for the solution of the initial value problem for the ordinary differential equation in \texttt{Minlog}.
We performed the program extraction in the exact real number arithmetic and showed how the extracted program can be optimized in terms of efficiency and maintaining a certain high level of abstraction.
The presented results might be fruitful for the reasoning on higher-order numerical schemes like Runge-Kutta.
 Furthermore the notion of system trajectory plays a crucial role in the stability of the equilibrium of nonlinear systems where the notion of uniform asymptotic stability is of a separate interest, due to the arising of special initial value problem from the comparison principle \cite{Khalil1996-nonlin-sys}. Thus the presented concepts may also be viewed as a part of preliminary work for the development of formalized nonlinear control theory.

\bibliography{bib/sos,bib/qe,bib/constr-math,bib/formal_math,bib/form-ver-ctrl,bib/minlog,bib/computable,bib/russian-school,bib/schwichtenberg,bib/computable-functionals,bib/control-theory,bib/cas,bib/dyn-sys,bib/analysis,bib/numerics}{}     
\bibliographystyle{IEEEtranN}                                                 		     

\end{document}